\documentclass[11pt]{article}
\usepackage{amsmath,amssymb}
\usepackage{epsfig}
\usepackage{sint_modi,cite}

\let\a=\alpha      \let\g=\gamma   
\let\e=\epsilon         
      \let\l=\lambda  \let\m=\mu
                 
\let\s=\sigma        

\newcommand{\be}{\begin{equation}}
\newcommand{\ee}{\end{equation}}
\newcommand{\bea}{\begin{eqnarray}}
\newcommand{\eea}{\end{eqnarray}}
\newcommand{\ba}{\begin{array}}
\newcommand{\ea}{\end{array}}
\def\bz{{\bar z}}
\def\tr{{\rm tr}}
\def\SUN{{\rm SU}(N)}
\def\sun{{\rm su}(N)}
\def\hM{\hat M}

\newcommand{\eq}[1]{Eq.~(\ref{#1})}
\newcommand{\fig}[1]{Fig.~\ref{#1}}

\def\Oto{\stackrel{\Omega}{\longrightarrow}}
\def\SO{S^{\rm orb}_{\rm W}}
\def\TA{T_{\rm A}}
\def\xp{x^\prime}
\def\Ap{{A^\prime}}
\def\g0p{g_0^\prime}
\def\Fp{{F^\prime}}
\def\thintablerule{\hrule height0.4pt}
\begin{document}

\rightline{HU-EP-04/65}

\vskip 1.5cm
\centerline{\LARGE Non--perturbative definition of five--dimensional}
\vskip 0.3 true cm
\centerline{\LARGE gauge theories on the $\mathbb{R}^4\times S^1/\mathbb{Z}_2$
orbifold}
\vskip 0.6 true cm
\centerline{\large Nikos Irges}
\vskip1ex
\centerline{\it High Energy and Elementary Particle Physics Division,}
\centerline{\it Department of Physics, University of Crete, 71003 Heraklion, Greece}
\centerline{\it E-Mail: {\tt irges@physics.uoc.gr}}
\vskip 0.3 true cm
\centerline{\large Francesco Knechtli}
\vskip1ex
\centerline{\it Institut f\"ur Physik, Humboldt Universit\"at,}
\centerline{\it Newtonstr. 15, 12489 Berlin, Germany}
\centerline{\it E-mail: {\tt knechtli@physik.hu-berlin.de}}
\vskip 0.8 true cm
\thintablerule
\vskip 2.0ex
\leftline{\bf Abstract}
\vskip 1.0ex\noindent
We construct a $\mathbb{Z}_2$ orbifold projection of $\SUN$ gauge theories
formulated in five dimensions with a compact fifth dimension.
We show through a non--perturbative argument that no boundary mass term
for the Higgs field, identified with some of the fifth dimensional
components of the gauge field, is generated, which would be quadratically
divergent in the five--dimensional ultraviolet cutoff.
This opens the possibility of studying these theories non--perturbatively in order
to establish if they can be used as effective weakly interacting theories at low
energies. We make preparations for a study on the lattice. In particular we show
that only Dirichlet boundary conditions are needed, which specify the breaking
pattern of the gauge group at the orbifold fixpoints.
\vskip 2.0ex
\thintablerule

\vskip-0.2cm

\section{Introduction}
\label{introduction}

Gauge theories with
extra space--like dimensions have attracted interest during the last few years. 
Even though departing from four dimensions leads into the ``wild'' domain
of non--renormalizable theories, there are perhaps 
reasons they should not be discarded immediately:
an ultraviolet (UV) cutoff $\Lambda$ (like the inverse lattice spacing)
can be introduced and the theory can be treated as an effective low--energy theory.
One is however not guaranteed that this 
is a consistent program unless
there exists a range of the cutoff $\Lambda$ where the
low--energy physical properties depend only weakly on $\Lambda$
(this is called the scaling region) and the theory is weakly interacting.
If this is the case then these theories could provide a solid starting point basis
for constructing non--supersymmetric extensions of
known and well tested physical theories.

We discuss more in detail this point for $\SUN$ gauge theories. Gauge invariance
guides the construction of the theory. We assume that the effective gauge
theory in $d$ Euclidean dimensions can be written as
\begin{eqnarray}
 S & = & \frac{1}{4}\int{\rm d}^dx\left[
 b\,F_{MN}^AF_{MN}^A +
 c\,D_LF_{MN}^AD_LF_{MN}^A
 \right] \,. \label{effa}
\end{eqnarray}
All quantities appearing in the above are dimensionless,
for example the action $S$ can be discretized on a lattice
and all quantities are pure numbers in units of the lattice spacing.
We neglect in \eq{effa}
other terms allowed by gauge invariance for reasons which will become
clear later. The index $A$ is the adjoint gauge group index, the indices $M,N,L$
are the Euclidean indices and $D_L$ is the gauge covariant derivative. The field
strength components $F_{MN}^A$ contain the bare gauge coupling $g_0$. 
The theory is defined in terms of the parameters $b$ and $c$.
We rescale coordinates, gauge field $A$ and
coupling as\footnote{
We would like to thank Jean Zinn--Justin for pointing out this to us.}
\begin{eqnarray}
 x & = & \Lambda \xp \,, \\
 A_M^A(x) & = & \zeta \Ap_M^A(\xp) \,, \\
 g_0 & = & \xi \g0p \,. \label{rengcoupling}
\end{eqnarray}
$\Lambda$ can be thought of a (large) momentum cutoff giving a dimension
to the physical primed (rescaled) quantities. The action \eq{effa} can be
interpreted as a low--energy effective theory for energies $E\ll\Lambda$.
The coupling $\g0p$ is the effective coupling at the scale $\Lambda$,
renormalized up to slowly varying renormalization factors. 
Requiring that the kinetic term
for $\Ap$ has the standard coefficient $1/4$ fixes
\begin{eqnarray}
 \zeta \,=\, b^{-1/2} \Lambda^{(2-d)/2} & \mbox{and} &
 \xi \,=\, b^{1/2} \Lambda^{(d-4)/2} \,.
\end{eqnarray}
In terms of the rescaled quantities the action \eq{effa} becomes
\begin{eqnarray}
 S & = & \frac{1}{4}\int{\rm d}^d\xp\left[
 \Fp_{MN}^A\Fp_{MN}^A +
 \frac{c}{b} \frac{1}{\Lambda^2}\,D^\prime_L\Fp_{MN}^AD^\prime_L\Fp_{MN}^A
 \right] \,. \label{effaresc}
\end{eqnarray}
We observe that the term with two covariant derivatives gives contributions
suppressed by a factor $(E/\Lambda)^2$. Even more suppressed are further terms
containing more fields and their derivatives. 
The action \eq{effaresc} is reminiscent of the Symanzik effective action
for renormalizable lattice theories
\cite{Symanzik:1981hc,Symanzik:1983dc,Symanzik:1983gh,Luscher:1998pe}.
The low--energy effective theory, for energies $E\ll\Lambda=1/a$ where $a$ is the
lattice spacing, is a continuum theory where the lattice spacing is the
expansion parameter appearing in positive powers.

In $d=5$ dimensions pure $\SUN$ gauge theories have at least two phases in
infinite volume. On a lattice when the bare gauge coupling $g_0$ is very large
the theory is confining, while at very small $g_0$ the theory is in a Coulomb
phase with massless gluons \cite{Creutz:1979dw,Beard:1997ic}. 
There must therefore be at least one phase--transition point between these phases.
An attempt to approach the continuum limit in the Coulomb phase
is by keeping the renormalized gauge coupling $\g0p$
fixed and increasing the UV cutoff $\Lambda$. From \eq{rengcoupling},
$\g0p = (b\Lambda)^{-1/2}g_0$ which requires the bare gauge coupling $g_0$ to be
scaled to larger values roughly proportionally to the square root of $\Lambda$.
Eventually the phase transition point is reached which implies that at
fixed renormalized coupling there is a maximal value of the cutoff and thus the
theory is trivial.

Triviality is a property shared also by some renormalizable theories, like
$\phi^4$ theory in four dimensions. Studies of this theory
formulated on a lattice provided strong evidence of the 
triviality of the continuum limit \cite{Luscher:1987ay,Luscher:1988ek}, 
i.e. removing the ultraviolet cutoff $\Lambda$ leads to
a zero renormalized gauge coupling. Nevertheless appreciable interactions
are observed in a scaling region for finite values of the lattice spacing.
An important difference is that the renormalized gauge coupling in the
$\phi^4$ theory approaches zero as the inverse logarithm of $a$.
In non--renormalizable theories the renormalized coupling approaches zero
with a power of $a$.

There exist alternative approaches to construct an 
effective theory for five--dimensional gauge theories.
In the $D$--theory regularization of field theory,
a five--dimensional gauge theory on the lattice arises as low--energy effective
theory of quantum link models \cite{Chandrasekharan:1996ih,Brower:1997ha}.
It has been shown that dimensional reduction to a four--dimensional $\SUN$
gauge theory can occur \cite{Schlittgen:2000xg}. The gauge coupling constant $g$
of the dimensionally reduced theory is given by $g^2\,=\,{\g0p}^2/\beta$, where 
$\g0p$ is the dimensionful gauge coupling of the
five--dimensional quantum link model
and $\beta$ is the extension of the fifth dimension.
Yet another approach to five--dimensional gauge theories is the investigation of
a non--perturbative UV fixed point.
In case such a fixed point exists, the limit of infinite UV cutoff $\Lambda$
could be taken. Its existence is suggested by
the epsilon expansion \cite{Gies:2003ic,Morris:2004mg} but it has
been elusive so far in lattice simulations.

In extra dimensional theories typically one assumes that the extra dimensions are 
compactified on some manifold, a torus in the simplest case.
The minimal of the clearly large number of possibilities is a gauge theory
with a single compact extra dimension.
The advantage of such a simple model is that one can make considerable
progress in understanding its quantum properties which 
becomes increasingly hard as the number 
of the extra dimensions grows or when the
extra dimensional theory is coupled to gravity. 
The UV behavior of the compact theory is 
the same as that of the uncompactified theory
so all the above comments and questions apply to it.
Nevertheless, compactification is well motivated by phenomenology.
The first phenomenological applications of large (${\rm TeV}^{-1}$)
extra dimensions to the Standard Model were proposed in
\cite{Antoniadis:1988jn, Antoniadis:1990ew}.
A striking example is that if the compact dimension is as large as
$1\,{\rm TeV}^{-1}$ then the electroweak symmetry breaking could proceed
by the Hosotani mechanism \cite{Hosotani:1983xw,Hosotani:1989bm}
without supersymmetry and avoiding the hierarchy problem.
The idea is to identify the Standard Model Higgs field with an extra dimensional
component of the gauge field for which
a non-trivial effective potential is conjectured.
Results at 1-loop 
\cite{Antoniadis:1993jp,Hatanaka:1998yp,Antoniadis:2001cv,Kubo:2001zc} 
support the viability of this scenario but only a
non-perturbative computation can prove its true validity.
Extra dimensions in connection with further
alternatives to the Higgs mechanism have received attention from recent
lattice studies \cite{Laine:2002rh}.
Also the dimensional reduction and localization of gauge fields have been
studied on the lattice in a three--dimensional model \cite{Laine:2004ji}.

Compactification introduces a new scale in the theory,
in the case of circle compactifications the radius $R$ of the circle $S^1$.
Like in the case of field theories at nonzero temperature \cite{ZinnJustin-Book}
dimensional reduction to four dimensions can be investigated.
The regime where the effective four--dimensional
theory can behave like a
weakly interacting field theory at low energies $E$ is now
\bea
 E \,\ll\, \frac{1}{R} \,\ll\, \Lambda \,. \label{range}
\eea
The four--dimensional theory
is effectively a theory of the Kaluza--Klein zero modes
of the five--dimensional fields. To understand this better recall that 
in the perturbative approach one typically fixes the gauge by imposing that the
gauge field is periodic in the compact coordinate. Then a Kaluza-Klein
expansion of the fields is possible and perturbative
calculations can be performed in a four--dimensional language, 
where one has four--dimensional
massless fields together with infinite towers of massive fields.
The breaking of the five--dimensional gauge invariance can be 
interpreted as a Higgs mechanism where the non--zero modes of the 
four--dimensional part of the gauge field absorb the non--zero modes of the
fifth dimensional component of the gauge field \cite{Karlhede:1982xi}. 
In perturbative calculations, after summing over the massive 
states using certain infinite sum regularization
methods, one can arrive at interesting results,  
such as the 1--loop mass of the adjoint Kaluza--Klein scalar 
(the Higgs field in this approach), which is found to be finite.  
It would be interesting to verify that this result 
is not just a gauge artifact, result of the specific gauge fixing method.

An obvious practical problem with a five--dimensional gauge theory 
(intended to be used for four--dimensional physics) is 
how to take a four--dimensional slice of it in such a way that this slice
resembles the physics that we observe. 
A possible solution to this problem turns out to be the same as
the solution to the problem of the non--existence of chiral fermions
in five dimensions. 
By changing the compact space from a circle $S^1$ of radius $R$
parametrized by the coordinate $x_5$
into an interval $S^1/\mathbb{Z}_2$ by the identification $x_5\longrightarrow -x_5$,
one obtains naturally four dimensional boundaries at the two ends of the interval
(which are just the fixed points of the projection)
with chiral fermions localized on them. 
The new space obtained in this way is called an orbifold.
One can embed the orbifold projection in a gauge theory by 
imposing certain boundary conditions on the gauge fields.
The orbifold projection can thus reduce
the gauge symmetry at the four--dimensional fixed points of the orbifold.
As a result, the Higgs field does not transform in the adjoint representation of the
gauge group as in the $S^1$ compactification but in some lower dimensional
representation, a property shared also by the Standard Model Higgs field. 
For recent promising phenomenological applications where such theories
are used to construct models for extensions of the Standard Model, see
Refs. \cite{Scrucca:2003ra,Biggio:2003kp,Martinelli:2005ix}.

The introduction of a fifth dimension 
in connection with chiral fermions on the lattice is at the basis of
the domain wall fermion formulation \cite{Kaplan:1992bt}.
It is also known that in the domain wall construction of chiral fermions the
domain wall can be replaced by a boundary through Dirichlet boundary conditions
\cite{Luscher:2000hn} which is precisely what one has in the orbifold construction.
The derivation of light four--dimensional fermions from a five--dimensional theory
with boundaries may be a concrete hint of the physical reality of compact extra
dimensions \cite{Chandrasekharan:2004cn}.
Stimulating progress related to the fermions comes 
from a recent work where
the orbifold construction has been used to formulate in four dimensions
lattice chiral fermions with Schr{\"o}dinger functional boundary conditions
\cite{Taniguchi:2004gf}.

In the orbifold compactification of the gauge theory 
there is a new problem that appears due to the presence of the boundaries:
fields acquire Dirichlet or Neumann boundary conditions at the fixpoints of the orbifold.
The formulation of a field theory with prescribed boundary values for some
of the field components requires in general additional renormalization. 
This has been first studied for renormalizable theories by
Symanzik \cite{Symanzik:1981wd,Luscher:1985iu}. 
There it was found that the presence of boundaries introduces
additional divergences and these induce boundary counterterms
with renormalization factors calculable in perturbation theory.
The important lesson therefore is that renormalization requires
counterterms localized on the {\em boundaries} of the theory and
this applies also to non--renormalizable theories in the parameter
range \eq{range}.
The renormalization pattern of a five--dimensional Yukawa theory
formulated on the ${\cal M}^4\times S^1/\mathbb{Z}_2$ orbifold
has been first discussed in \cite{Georgi:2000ks}. There, counterterms
localized on the boundaries and logarithmically divergent in the cutoff
have been computed in perturbation theory at 1--loop order. 
Five--dimensional gauge theories formulated on 
${\cal M}^4\times S^1/\mathbb{Z}_2$ have been considered in
\cite{Kawamura:1999nj,Contino:2001si,vonGersdorff:2002as,Cheng:2002iz,vonGersdorff:2005ce}.
The main result of \cite{vonGersdorff:2002as} was that
at 1--loop level a {\em boundary mass counterterm} for the Higgs field
(identified with some of the five--dimensional components of the gauge field)
is absent. This term would represent a correction to the Higgs mass squared
proportional to ${\g0p}^2\Lambda^2/R$ for the zero modes of the four--dimensional 
low energy theory defined at the boundaries, introducing a hierarchy problem.
It was not clear though if a boundary counterterm is absent also at
higher orders in perturbation theory. A strong indication for this is
the shift symmetry argument given in
Refs. \cite{vonGersdorff:2002rg,vonGersdorff:2002us}.

The scenario of dimensional reduction of five--dimensional orbifold gauge theories
to four--dimensional theories of gauge and Higgs fields at the orbifold fixpoints
is supported by perturbative calculations. The mass of the Higgs field is generated
through radiative corrections. At 1--loop it is found independent of the
five--dimensional UV cutoff $\Lambda$. However,
since the theory is non--renormalizable, a sensitivity to $\Lambda$
is expected at higher orders in perturbation theory. The results of Ref.
\cite{vonGersdorff:2005ce} show that at 2--loop order the Higgs mass
receives a contribution logarithmic in $\Lambda$, generated by insertions
of boundary terms in finite 1--loop bulk graphs.
It is not clear to us whether radiative corrections will generate
power divergent contributions at even higher orders in perturbation theory.

In this paper we make preparations to study on the lattice $\SUN$ pure
gauge theories on the orbifold $\mathbb{R}^4\times S^1/\mathbb{Z}_2$.
The idea we would like to investigate is if in principle 
one could have a four--dimensional non--supersymmetric effective theory 
coupled to a Higgs field without a hierarchy problem.
In section \ref{orbifold} and \ref{boundaries}
we present a proof that a boundary mass counterterm for the Higgs field is absent.
Regarding boundary counterterms,
in a non--perturbative formulation the main problem turns out to be to
develop a gauge invariant method for their classification, despite the fact 
that gauge invariance may be broken at the boundaries.
The basic tool for developing such a method is 
the introduction of a spurion field which restores the
gauge invariance of the theory broken by the orbifold boundaries.
In section \ref{lattice}
we construct the lattice orbifold theory.
In section \ref{conclusion}
we make a short summary of this work.              
\section{The orbifold}
\label{orbifold}

The orbifold projection identifies field components under the transformations
of a discrete symmetry group $K$. Here we consider five--dimensional gauge theories with gauge group
${G}=\SUN$ and $K=\mathbb{Z}_2$ formulated in Euclidean space.
We use capital Latin letters $M,N,\ldots \,=\, 0,1,2,3,5$ to denote the
five--dimensional Euclidean index and small Greek letters
$\mu,\nu,\ldots \,=\, 0,1,2,3$ to denote the four--dimensional part.
For the coordinates we will use the shorthand notation
$z=(x_{\mu},x_5)$ and $\bz=(x_{\mu},-x_5)$. 
In the following we will suppress $x_{\m}$ whenever its
explicit appearance is not necessary.

We introduce the $\mathbb{Z}_2$ reflection ${\cal R}$ in the fifth dimension
\bea
{\cal R}\,z & = & \bz \,. \label{Rtranscoor}
\eea
Next, we define the $\mathbb{Z}_2$ reflection on a rank--$r$ tensor field $C(z)$ as
\bea
 \left({\cal R}\,C_{M_1M_2 \cdots M_r}\right)(z) =
 \a_{M_1}\a_{M_2}\cdots \a_{M_r}\,C_{M_1M_2 \cdots M_r}({\cal R}\,z) \,,
 \label{Rtrans}
\eea
where no sum on the $M_i$ is implied on the right hand side.
The intrinsic parities $\a_M$ are defined by $\a_{\mu}=1$ and $\a_5=-1$.
Since tensor fields can be obtained through derivatives of fields, the relation
\bea
 \left({\cal R}\,\partial_MC_{M_1M_2 \cdots M_r}\right)(z) & = &
 \a_M\a_{M_1}\a_{M_2}\cdots \a_{M_r}
 \left(\partial_MC_{M_1M_2 \cdots M_r}\right)({\cal R}\,z) \,,
 \label{Rtransder}
\eea
holds. Incidentally this implies that ${\cal R}$ and the derivative operator
$\partial_M$ commute
\bea
 [{\cal R}\,,\partial_M] & = & 0 \,.
\eea
Also \eq{Rtrans} is consistent with the property
\bea
 ({\cal R}\,C \cdot D)(z) & = & ({\cal R}\,C)(z) \cdot ({\cal R}\,D)(z)
\eea
for the product of any two tensor fields $C(z)$ and $D(z)$. In the following
for ${\cal R}$ and similarly for all the other operators we will write
${\cal R}\,C(z)$ as a shorthand for $({\cal R}\,C)(z)$.

Inspired by the geometric description of gauge fields on 
$R^4\times S^1/\mathbb{Z}_2$ (see Appendix \ref{app_circle})
one can formulate the orbifold theory on the strip
\bea
 I_0 & = & \{x_\mu,0\leq x_5 \leq \pi R\} \, \label{strip}
\eea
without reference to the circle. The following construction yields the proper
boundary conditions on the boundary planes at $x_5=0$ and $x_5=\pi R$.

One starts with an $SU(N)$ gauge theory defined on the open set 
$I_{\e}=\{x_\mu,x_5\in(-\e , \pi R + \e)\}$
with a gauge field $A_M (z)$ defined everywhere on $I_{\e}$ and a
{\em spurion field} \footnote{
We would like to thank Martin L{\"u}scher for suggesting this to us.}
${\cal G}(z)\in\SUN$ defined in the neighborhoods
$O_1 \,=\, \{x_\mu,x_5\in(-\e ,\e )\}$  and
$O_2 \,=\, \{x_\mu,x_5\in(\pi R - \e ,\pi R + \e)\}$
that satisfies 
\bea
({\cal R}\,{\cal G})\,{\cal G} & = & \pm 1 \,, \label{RG}
\eea
with ${\cal R}$ the reflection operator. At the fixpoints $x_5=0$ and $x_5=\pi R$
of ${\cal R}$, \eq{RG} states that ${\cal G}^2=1$.
The gauge field on $O_i$ is constrained by
\bea
{\cal R}\,A_M & = & {\cal G}\,A_M\,{\cal G}^{-1} + 
{\cal G}\, \partial_M{\cal G}^{-1} \,, \label{RA}
\eea
which implies ${\cal R}\,F_{MN}\,=\,{\cal G}\,F_{MN}\,{\cal G}^{-1}$.
In Appendix \ref{app_circle}
it is shown that the spurion field can be identified with a
transition function that is required when defining gauge fields on the circle
using two charts. The property \eq{RG} expresses the gluing condition of the
two charts.
The transformation property of the spurion field under a gauge transformation
is such that the constraint \eq{RA} is covariant under
gauge transformations $\Omega\in\SUN$. This is the case for
\bea
{\cal G} & \Oto & ({\cal R}\,\Omega)\,{\cal G}\,\Omega^{-1} \,. \label{gtG}
\eea
The covariant derivative of ${\cal G}$ can be defined
on the neighborhoods $O_i$ by requiring that it transforms like ${\cal G}$.
Such a covariant derivative is
\bea
 D_M{\cal G} & = & \partial_M{\cal G} + 
                   ({\cal R}\,A_M)\,{\cal G} - {\cal G}\,A_M \,
\eea
and in fact, \eq{RA} implies that 
\bea
D_M\, {\cal G}\equiv \, 0. \label{covder}
\eea
By means of the constraints \eq{RG}, \eq{RA} and
\bea
 {\cal R}\,\Omega & = & \Omega \label{RO}
\eea
which ensures that all gauge transformations on $I_{\e}$ are local,
the orbifold theory can be consistently defined on the strip $I_0$
respecting the $\SUN$ gauge symmetry.

For any $\e\neq0$ the theories are gauge invariant and equivalent.
The breaking of the gauge symmetry is realized by taking the limit
$\epsilon \longrightarrow 0$. In this limit the neighborhoods $O_i$ shrink
to single points and one is left with boundaries at $x_5=0$ and $x_5=\pi R$.
We approach the limit $\epsilon \longrightarrow 0$ so that (in the limit),
the spurion field and its derivatives take the value
\begin{eqnarray}
 {\cal G}(0) \,=\, {\cal G}(\pi R) & = & g \,, \label{bcG} \\
 \partial_5^p{\cal G}(0) \,=\, \partial_5^p{\cal G}(\pi R) & = & 0 \,,
 \quad p\in\mathbb{N}\,,\; p>0 \label{bcdG}
\end{eqnarray}
for a constant matrix\footnote{
One could in principle take a different matrix $g$ for $x_5=0$ and $x_5=\pi R$.}
$g$ obeying $g^2=\pm 1$ by virtue of \eq{RG}.
We will specify the matrix $g$ below.
Since $g$ is constant all derivatives $\partial_\mu$ of ${\cal G}$ vanish
as $\epsilon \longrightarrow 0$.
From \eq{gtG} it is immediately clear that only gauge transformations
for which
\begin{eqnarray}
 \Omega & = & g\,\Omega\,g 
 \qquad\mbox{at}\;\; x_5=0\;\; \mbox{and}\;\; x_5=\pi R \label{resgt}
\end{eqnarray}
are still a symmetry of the theory. Taking the limit $\epsilon \longrightarrow 0$
in \eq{RA} yields the Dirichlet boundary conditions
\begin{eqnarray}
 \alpha_M\,A_M & = & g\,A_M\,g
 \qquad\mbox{at}\;\; x_5=0\;\; \mbox{and}\;\; x_5=\pi R \,, \label{DbcA}
\end{eqnarray}
where no sum on $M$ is implied on the left hand side.

We now have a prescription to obtain the correct boundary conditions for {\em any}
field derived from $A_M$. One starts in the gauge invariant theory ($\e\neq0$)
where the field $A_M$ is constrained by \eq{RA}. Then the limit
$\e \longrightarrow 0$ is taken using the properties of ${\cal G}$ in
\eq{bcG} and \eq{bcdG}. For example we obtain
the following Neumann boundary conditions
\bea
 -\a_M\,\partial_5A_M & = & g\,\partial_5A_M\,g
 \qquad\mbox{at}\;\; x_5=0\;\; \mbox{and}\;\; x_5=\pi R \,, \label{NbcA}
\eea
where no sum on $M$ is implied on the left hand side.
From \eq{RA} Dirichlet boundary conditions follow for the field strength
tensor
\bea
 \a_M\,\a_N\,F_{MN} & = & g\,F_{MN}\,g
 \qquad\mbox{at}\;\; x_5=0\;\; \mbox{and}\;\; x_5=\pi R \,, \label{DbcF}
\eea
where no sum on $M$ and $N$ is implied on the left hand side. The point we
would like to emphasize here is that our construction provides {\em all} the
necessary boundary conditions that {\em define} the orbifold theory.

The gauge symmetry at the boundaries is broken to a subgroup $\mathcal{H}$
of $SU(N)$ by the group conjugation in \eq{resgt}.
The latter is an inner automorphism of the Lie algebra
and for $g$ one can take
\bea
 g \,=\, e^{-2\pi i V\cdot H} \,,\label{orbma}
\eea
with $H=\{H_i\}\,,\;i=1,\ldots,{\rm rank}(\SUN)=N-1$
the hermitian generators of the Cartan subalgebra of $\SUN$. $V=\{V_i\}$ is
a constant $(N-1)$--dimensional twist vector.
In general, an inner automorphism breaks the gauge group as
\bea
 G \,=\, SU(p+q) & \longrightarrow & 
 \mathcal{H} \,=\, SU(p)\times SU(q)\times U(1) \,.
 \label{brokengroup}
\eea
As shown in Appendix \ref{app_gauge} under a group conjugation by $g$ the
hermitian $\SUN$ generators $T^A$, $A=1,\ldots,N^2-1$ transform as
$g\,T^A\,g=\eta^A\,T^A$, where $\eta^A=\pm1$ is their parity. The
generators are divided into unbroken generators $T^a$ with $\eta^a=1$ and
broken generators $T^{\hat{a}}$ with $\eta^{\hat{a}}=-1$.
The above imply that in the adjoint representation the matrix elements of
the conjugation matrix are simply $g^{AA^\prime} = \eta^A\delta^{AA^\prime}$.
In terms of the gauge field components $A_M=-ig_0A_M^AT^A$ the boundary
conditions \eq{DbcA} and \eq{NbcA} read
\bea
 A_\mu^{\hat{a}} \,=\, 0 & \quad\mbox{and}\quad & A_5^a \,=\, 0 
 \qquad\quad\mbox{at}\;\; x_5=0\;\; \mbox{and}\;\; x_5=\pi R \,, \label{DbcAcomp} \\
 \partial_5A_\mu^a \,=\, 0 & \quad\mbox{and}\quad & \partial_5A_5^{\hat{a}} \,=\, 0
 \qquad\mbox{at}\;\; x_5=0\;\; \mbox{and}\;\; x_5=\pi R \,. \label{NbcAcomp}
\eea
In the Kaluza-Klein decomposition,
the zero modes of $A_5^{\hat{a}}$ are the Higgs fields
of the four--dimensional low--energy effective theory defined
at the orbifold boundaries.
The zero modes of $A_\mu^a$ are the gauge bosons,
which generate the residual gauge group $\mathcal{H}$. The Higgs fields
transform in some representation of $\mathcal{H}$.

The simplest possibility in \eq{brokengroup}
is the breaking pattern $SU(2)\longrightarrow U(1)$
which can be achieved with the twist vector $V=1/2$. This twist vector results in
$\{\eta^A\}=\{-1,-1,+1\}$ (which specifies $g$ in the adjoint representation)
and the gauge boson branching  
${\bf 3}={\bf 1}_1 + {\bf 1}_{-1}+{\bf 1}_0$, where the subscripts are the $U(1)$
charges. There are two charged Higgs scalars.
In the fundamental representation one can use $g=-i\s_3$.

The next simplest case is $SU(3)\longrightarrow SU(2)\times U(1)$.
The twist vector is $V=(0,\sqrt{3})$ and $\{\eta^A\}=\{+1,+1,+1,-1,-1,-1,-1,+1\}$.
The gauge bosons branch as ${\bf 8}={\bf 3}_0+{\bf 2}_1+{\bf 2}_{-1}+{\bf 1}_0$.
There are two Higgs fields in the {\em fundamental}
representation of the unbroken $SU(2)$ gauge group with $U(1)$ charges $\pm1$.
The matrix $g$ in the fundamental representation in this case is
$g={\rm diag}(-1,-1,+1)$.

\section{The boundary terms}
\label{boundaries}

In general the presence of boundaries in a field theory leads to new
divergences.
Symanzik studied the Schr{\"o}dinger functional for renormalizable theories
\cite{Symanzik:1981wd,Luscher:1985iu}.
The Schr{\"o}dinger functional is a formulation of
field theories with prescribed boundary values for some of the field components.
The expectation is that to make these theories finite
all the possible (i.e. consistent with the symmetries of the theory) counterterms
localized on the boundaries have to be added.
This expectation has been confirmed for the massless scalar $\phi^4$ theory
\cite{Symanzik:1981wd} and for QCD
\cite{Luscher:1992an,Sint:1993un,Sint:1995rb,Luscher:1996sc}.

The aim of our work is to define the orbifold theory on a Euclidean lattice. 
For renormalizable quantum field theories
the dependence on the lattice spacing 
can be described at low energies in terms of
a continuum local effective theory. The associated Symanzik effective action
\cite{Symanzik:1981hc,Symanzik:1983dc,Symanzik:1983gh,Luscher:1998pe}
includes bulk and boundary terms.
As explained in the Introduction the non--renormalizable
five--dimensional orbifold theory makes sense only as an
effective theory for energies much below a finite cutoff,
in our case given by the inverse lattice spacing $1/a$.
In this regime the theory behaves effectively like a continuum theory
and renormalized perturbation theory applies.
The difference with respect to renormalizable theories is that renormalization
requires at each order the subtraction in the effective action
of new divergent terms of increasing dimension.
We expect that the orbifold theory defined on the strip $I_0$ and put on a lattice
can be described by a continuum Symanzik effective action.

In Section \ref{orbifold} we have constructed the orbifold theory on the strip $I_0$
as a limit $\e\longrightarrow 0$ of a gauge invariant theory defined on the open
set $I_\e$. The latter theory has the full $\SUN$ gauge invariance. In particular
we showed that our construction provides the orbifold boundary conditions on $I_0$
for any field. 
The only dangerous terms {\em at tree level}
in the Symanzik effective action on $I_0$ are
{\em boundary terms} described by local composite fields
$\mathcal{O}_i(x)$ of dimension less than or equal to four
contributing with a boundary action
\bea
 \delta S_{\rm b}[A] & = & \int{\rm d}^4x \sum_i Z_i \left\{
 \mathcal{O}_i(x)|_{x_5=0} + \mathcal{O}_i(x)|_{x_5=\pi R} \right\} \,.
 \label{actionb}
\eea
The canonical dimensions of the operators $\mathcal{O}_i$ determine the superficial
degree of divergence with the cutoff of the renormalization constants $Z_i$.
The boundary counterterms are generated by terms in the Symanzik effective
action for the gauge invariant theory on $I_\e$ containing ${\cal G}$.
To make a list of all possible gauge invariant terms involving ${\cal G}$,
we first note that $\tr\{{\cal G}\}$, $\tr\{{\cal G}^2\}$, $\ldots$ contribute
only an irrelevant constant and there is {\em no} kinetic term for ${\cal G}$.
The lowest dimensional gauge invariant terms are therefore the dimension five terms
\bea
 \frac{1}{{\g0p}^2}{\rm Re}\,\tr\{{\cal G}\,F_{MN}\,F_{MN}\} & = & \frac{1}{2{\g0p}^2}\left(
 \tr\{{\cal G}\,F_{MN}\,F_{MN}\} + \tr\{{\cal G}^{-1}\,F_{MN}\,F_{MN}\}\right) \,,\nonumber\\
 \frac{1}{{\g0p}^2}{\rm Re}\,\tr\{{\cal G}\,F_{MN}\,{\cal G}\,F_{MN}\} & = & \frac{1}{2{\g0p}^2}\left(
 \tr\{{\cal G}\,F_{MN}\,{\cal G}\,F_{MN}\} + 
 \tr\{{\cal G}^{-1}\,F_{MN}\,{\cal G}^{-1}\,F_{MN}\} \right)\,.
\eea
They generate the boundary terms
\bea
 \frac{1}{{\g0p}^2}\tr\{g\,F_{MN}(z)\,F_{MN}(z)\} \,,\quad
 \frac{1}{{\g0p}^2}\tr\{g\,F_{MN}(z)\,g\,F_{MN}(z)\},
 \label{boundaryterms}
\eea
invariant under the residual gauge transformations $\Omega$ satisfying \eq{resgt}. 
Being of dimension five these terms give at tree level
a contribution to the Symanzik action proportional to the lattice spacing.
We have hence proven that no boundary terms proportional at tree level
to inverse powers of the lattice spacing exist for the orbifold theory.

The first term in \eq{boundaryterms} can be evaluated taking
the generators in the adjoint representation $(\TA^A)^{BC} = -if^{ABC}$.
Evaluating the trace, one has
\bea
 \tr\{g\,\TA^A\,\TA^{A^\prime}\} & = & f^{ABC}f^{A^\prime BC}\eta^C \,=\,
 \left(C_2({\cal H})-\frac{1}{2}C_2(G)\right)(\eta^A+1)\delta^{AA^{\prime}}
 \,,
\eea
where ${\cal H}$ is the unbroken gauge subgroup at the boundaries and thus
we get the boundary term
\bea
 -\left(C_2({\cal H})-\frac{1}{2}C_2(G)\right)
 F_{\mu\nu}^aF_{\mu\nu}^a \,.
 \label{bt2}
\eea
In the above $C_2(G)$ and $C_2({\cal H})$ 
is the quadratic Casimir invariant of the unbroken group $G$ and 
its subgroup ${\cal H}$ respectively.
In the fundamental representation $C_2(\SUN)=1/2$. 
Counterterms of the type \eq{bt2} were indeed encountered at 1--loop
in perturbation theory \cite{vonGersdorff:2002as,Cheng:2002iz} with
logarithmically divergent Z--factors.

The second term in \eq{boundaryterms} can be evaluated using
$\tr\{g\,T^A\,g\,T^{A^\prime}\} = C_2(G)\eta^A\delta^{AA^{\prime}}$
yielding the boundary term
\bea
 -\frac{1}{2}C_2(G)
 \{F_{\mu\nu}^aF_{\mu\nu}^a-F_{\mu5}^{\hat{a}}F_{\mu5}^{\hat{a}}\} \,.
 \label{bt1}
\eea
Terms of this type do not appear at 1--loop, however they are expected to arise
at 2--loops even though such perturbative computation has not been performed yet. 

Finally, notice that the term
\bea
 \tr\{[A_M(z),g][A_M(z),g]\} \,,\label{Higgsterm}
\eea
is invariant under \eq{resgt}.
Using the boundary conditions \eq{DbcAcomp} this term is equal to
$2{\g0p}^2A_5^{\hat a}A_5^{\hat a}$, a would be quadratically divergent
boundary mass term for the Higgs.
An operator of the $\e\ne 0$ effective action
that could give rise to such a term is
\bea
{\rm tr} \{ D_M {\cal G} \, D_M {\cal G} \} \,,\label{wouldbeHiggsterm}
\eea
which is however identically zero, by \eq{covder}.
In fact, it is not hard to check that {\em none} of the operators
of the $\e\ne 0$ effective action containing
the spurion field (such as those in \eq{boundaryterms})
can induce a boundary Higgs mass term. 

It would be tempting at this point to conclude that the Higgs mass
in the orbifold theory is non---perturbatively finite. 
Indeed, if dimensional reduction occurs due to the compactification of the
fifth dimension,
as it happens for $\SUN$ gauge theories at nonzero temperature
in four dimensions \cite{ZinnJustin-Book},
the Kaluza--Klein zero modes of the fields $A_5^{\hat{a}}$ play the role of
Higgs fields in the four--dimensional low--energy effective theory.
In the dimensionally reduced theory a bulk mass term for the Higgs is allowed.
The 1--loop perturbative prediction is
\cite{vonGersdorff:2002as,Cheng:2002iz}
\bea
 m_h^2 & = & \frac{3}{32\pi^4R^2}\frac{\g0p}{\pi R}\zeta(3)3C_2(G), \,
 \label{higgsmasspt}
\eea
a manifestly finite result ($\g0p$ is the renormalized gauge coupling).
At higher orders of perturbation theory
there can be mixing of bulk and boundary radiative effects.
For example, one can have a finite bulk 1--loop correction to the Higgs mass 
infected at 2--loop order by divergences due to
{\em insertion of boundary counterterms} like \eq{boundaryterms}
\cite{vonGersdorff:2002as}. 
The explicit 2--loop computation of Ref. \cite{vonGersdorff:2005ce}
has indeed found these effects.
At 2--loop order the Higgs mass is logarithmically sensitive to the cutoff.
At this point only a non--perturbative computation can establish if there is
a scaling region where these higher order corrections to the
1--loop prediction \eq{higgsmasspt} are negligible.

\section{Lattice formulation}
\label{lattice}

We consider now a Euclidean five--dimensional hypercubic lattice with
lattice spacing $a$. The points have coordinates
$z=a\,(n_0,n_1,n_2,n_3,n_5)$
with $n_\mu=0,1,\ldots,N_{\mu}-1 \,,\; \mu=0,1,2,3$    
and
$n_5=-N_5,-N_5+1,\ldots,N_5-1$.
The physical extensions of the lattice are             
$L_\mu=N_\mu\,a$ and
$2\pi R=2N_5\,a$.
The gauge variables on the lattice consist of the links $U(z,M)\in\SUN$,
which are the parallel transporters for $\SUN$ vectors from $z+a\hM$ to $z$
along the straight line connecting these two points ($\hM$ is the unit vector
in direction $M$). Under gauge transformations
\bea
 U(z,M) & \Oto & \Omega(z)\,U(z,M)\,\Omega^{\dagger}(z+a\hM) \,.
\eea
We impose periodic boundary conditions on the gauge field and on the
gauge transformations in all five directions.
As the gauge action we take the Wilson action
\bea
 S_{\rm W}[U] & = & \frac{\beta}{2N}\sum_p\tr\{1-U(p)\} \,, \label{waction}
\eea
where the sum runs over all oriented plaquettes $p$ on the lattice.
We set
\bea
 \beta & = & \frac{2N}{{\g0p}^2}a \label{bareglat}
\eea
and identify $\g0p$ as the bare dimensionful gauge coupling on the lattice.

\begin{figure}[t]
\centerline{\epsfig{file=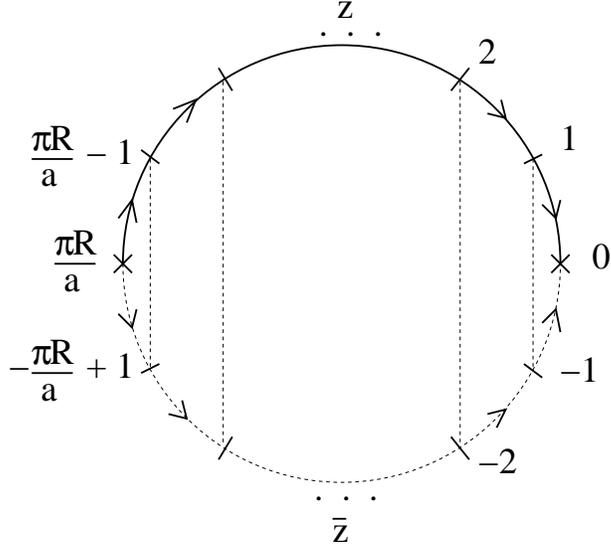,width=8cm}}
\caption{Representation of the $S^1/\mathbb{Z}_2$ orbifold projection in the
fifth dimension on the lattice. \label{f_orb}} 
\end{figure}

Unlike in perturbation theory in
the continuum, on the lattice the periodicity of the gauge links
does not break gauge invariance.
There is no need to introduce transition functions like we had to do in the
continuum formulation in Appendix \ref{app_circle}. The orbifold theory can
be therefore defined in a more straightforward way.
Given a continuum gauge field $A_M$ the gauge links can be reconstructed as
\bea
 U(z,M) & = & \mathcal{P}\exp\left\{a\int_0^1{\rm d}t\,A_M(z+a\hM-ta\hM)
              \right\} \,, \label{paralleltra}
\eea
where the symbol $\mathcal{P}$ implies a path ordered exponential such that
the fields at larger values of the integration variable $t$ stand to the left
of those with smaller $t$.
From the reflection ${\cal R}$, defined in \eq{Rtranscoor} and \eq{Rtrans},
and the group conjugation
\bea
 {\cal T}_g\,A_M & = & g\,A_M\,g \,, \label{Tgtrans}
\eea
with $g$ specified in \eq{orbma}, it is easy to derive the corresponding
$\mathbb{Z}_2$ transformations acting on the gauge links.
Under the reflection ${\cal R}$ the gauge links transform as
\bea
 {\cal R}\,U(z,\mu) \, = \, U(\bz,\mu)  &,&
 {\cal R}\,U(z,5) \, = \, U^{\dagger}(\bz-a\hat{5},5) \,. \label{refllat} 
\eea
\fig{f_orb} schematically represents the reflection ${\cal R}$ on the lattice.
%
%
Under the group conjugation ${\cal T}_g$ the gauge links transform as
\bea
 {\cal T}_g\,U(z,M) & = & g\,U(z,M)\,g \,. \label{twistglat}
\eea
The action \eq{waction} is invariant under the combined $\mathbb{Z}_2$
transformation $\Gamma={\cal R}\,{\cal T}_g$.
Consequently we embed the orbifold projection in the lattice theory through
\bea
 \frac{1-\Gamma}{2}\,U(z,M) & = & 0 \,, \label{orbiprojlat}
\eea
where $(1-\Gamma)/2$ is a projector.
For the gauge links in the four--dimensional planes defined by the
fixpoints $z=\bz$ of ${\cal R}$, \eq{orbiprojlat} implies
\bea
 U(z,\mu) & = & g\,U(z,\mu)\,g
 \qquad\mbox{at}\;\; n_5=0\;\; \mbox{and}\;\; n_5=N_5=\pi R/a \,. \label{Dbclat}
\eea
These constraints break the gauge group $\SUN$ down to the
subgroup $\cal{H}$ \eq{brokengroup} depending on the choice of $g$.
As discussed in Section \ref{orbifold}, the generators $T^a$ of
$\cal{H}$ satisfy $[T^a,g]=0$.

The lattice orbifold theory can now be defined on the strip
$I_0=\{n_\mu,0\le n_5\le N_5\}$. The action is
\bea
 \SO[U] & = & \frac{\beta}{2N}\sum_pw(p)\,\tr\{1-U(p)\} \label{alat}
\eea
where the sum runs now over all oriented plaquettes in the strip.
The weight $w(p)$ is
\bea
 w(p) & = & \left\{\begin{array}{ll}
                   \frac{1}{2} & \mbox{if $p$ is a plaquette in the
                   ($\mu$\,$\nu$)--planes at $n_5=0$ and $n_5=N_5$,} \\
                   1 & \mbox{in all other cases.}
                   \end{array} \right.
\eea
The Dirichlet boundary conditions are specified by \eq{Dbclat}. 
The normalization $\beta/(2N) = a/{\g0p}^2$ in \eq{alat} is such that the
continuum action on the strip is reproduced in the naive continuum limit.
The theory is invariant under gauge transformations
\bea
 \Omega(z) & \in & \left\{\begin{array}{ll}
                          \cal{H} & 
                          \mbox{at the boundary planes $n_5=0$ and $n_5=N_5$,} \\
                          \SUN    & \mbox{otherwise.}
                          \end{array} \right. \label{latogt}
\eea
We are left to prove that in the continuum limit 
the lattice orbifold projection \eq{orbiprojlat}
reproduces the boundary conditions \eq{DbcA} and \eq{NbcA}.
To this end, a gauge field on the lattice can be introduced through
$U(z,M)=\exp\{aA_M(z)\}$.

In the classical continuum limit, i.e. expanding
\eq{orbiprojlat} in powers of the lattice spacing $a$,
we get at the fixpoints
\bea
 A_\mu(z) & = & g\,A_\mu(\bz)\,g + \mathcal{O}(a) \,, \label{DbcAlat} \\
 A_5(z)   & = & -g\,A_5(\bz)\,g  + \mathcal{O}(a) \,. \label{NbcAlat}
\eea
The leading term at $z=\bz$
gives the Dirichlet boundary conditions \eq{DbcA}.

At the quantum level the gluon propagator (not the vertices)
carries the information about the boundaries.
The propagator on the lattice can be constructed
extending a trick used in Refs. \cite{Georgi:2000ks,Puchwein:2003jq}.
We observe that the orbifold constraint \eq{orbiprojlat} is satisfied automatically
by the gauge links
\bea
 U_{\Gamma}(z,M) & = & \exp\left\{a\frac{1+\Gamma}{2}B_M(z)\right\} \,,
 \label{Utrick}
\eea
where $B_M(z)$ is an unconstrained gauge field on the full periodic lattice.
It is easy to check that $U_{\Gamma}(z,M)\in\SUN$ and in particular that
at the fixpoints $z=\bz$ of ${\cal R}$ the links $U_{\Gamma}$ are elements of
its subgroup ${\cal H}$, as expected.
We use the hermitian basis of generators $T^A$ and under group conjugation
$g\,T^A\,g=\eta^A\,T^A$ (no sum on $A$), see Appendix \ref{app_gauge}.
The gauge field components are $A_M=-i\g0p A_M^CT^C$.
The building block is the propagator on a five--dimensional periodic
lattice
\bea
 \Delta_{MM^{\prime}}^{CC^{\prime}}(z-z^{\prime}) & = &
 \left(\prod_{\mu}N_{\mu}N_5\right)^{-1}\sum_p
 {\rm e}^{ip(z-z^{\prime})}{\rm e}^{ia(p_M-p_{M^{\prime}})/2}
 \tilde{\Delta}_{MM^{\prime}}^{CC^{\prime}}(p) \,, \label{latperpropcoor} \\
 \tilde{\Delta}_{MM^{\prime}}^{CC^{\prime}}(p) & = &
 \delta_{CC^{\prime}} \left\{  \frac{\delta_{MM^{\prime}}}{{\hat p}^2}
 -(1-\xi)\frac{{\hat p}_M{\hat p}_{M^{\prime}}}{{\hat p}^4}\right\} \,,
 \label{latperpropmom}
\eea
where 
${\hat p}_M = (2/a)\sin(ap_M/2)$
and the sum in \eq{latperpropcoor} runs over the momenta in the Brillouin zones
$p_{\mu} = 2\pi n_\mu/L_{\mu}$ and $p_5 = n_5/R$ with
$n_\mu=0,1,\ldots,N_{\mu}-1 \,,\; \mu=0,1,2,3$ and
$n_5=-N_5,-N_5+1,\ldots,N_5-1$.
Note that the gauge field on the lattice is naturally associated with the midpoint
of the link. In \eq{latperpropmom} we use the Lorentz gauge 
with parameter $\xi$ in the gauge fixing term.
The propagator on the orbifold is
\bea
 ({\Delta^{\rm orb}})_{MM^{\prime}}^{CC^{\prime}}(z,z^{\prime}) & = &
 \frac{1}{2}\left\{\Delta_{MM^{\prime}}^{CC^{\prime}}(z-z^{\prime}) +
 \alpha_{M^{\prime}}\,\eta^C\,\Delta_{MM^{\prime}}^{CC^{\prime}}(z-\overline{z^{\prime}})\right\}
 \,.\label{latticepropcoor}
\eea
To check that the correct Neumann boundary conditions are obtained in the
continuum limit of the lattice theory we use for example the identities
\bea
 ({\Delta^{\rm orb}})_{5\mu}^{CC^{\prime}}(\bz-a\hat{5},z^\prime) & = &
 -\eta^C\,({\Delta^{\rm orb}})_{5\mu}^{CC^{\prime}}(z,z^\prime) \,, \\
 ({\Delta^{\rm orb}})_{55}^{CC^{\prime}}(\bz-a\hat{5},z^\prime) & = &
 -\eta^C\,({\Delta^{\rm orb}})_{55}^{CC^{\prime}}(z+a\hat{5},z^\prime) \,.
\eea
Setting $C=\hat{a}$ and using the lattice forward $\partial$ and backward
$\partial^*$ derivatives, yields
\bea
 \partial_5^*({\Delta^{\rm orb}})_{5\mu}^{\hat{a}C^{\prime}}(z,z^{\prime})|_{z=\bz}
 & = & 0 \,, \label{latNbc5mu} \\
 (\partial_5+\partial_5^*)
 ({\Delta^{\rm orb}})_{55}^{\hat{a}C^{\prime}}(z,z^{\prime})|_{z=\bz} & = & 0 \,,
 \label{latNbc55}
\eea
which give in the continuum limit $\partial_5A_5^{\hat{a}}=0$ at $z=\bz$. Similarly
we obtain
\bea
 (\partial_5+\partial_5^*)
 ({\Delta^{\rm orb}})_{\mu\mu^\prime}^{aC^{\prime}}(z,z^{\prime})|_{z=\bz}
 & = & 0 \,, \label{latNbcmumu} \\
 \partial_5({\Delta^{\rm orb}})_{\mu5}^{aC^{\prime}}(z,z^{\prime})|_{z=\bz} & = & 0
 \,, \label{latNbcmu5}
\eea
which give in the continuum limit $\partial_5A_\mu^{a}=0$ at $z=\bz$.
We have therefore proven that the lattice orbifold propagator carries
in the continuum limit the Neumann boundary conditions \eq{NbcAcomp}.

Finally, a brief but important comment about the Higgs mass.
It will be certainly very interesting to compare 
the perturbative result \eq{higgsmasspt} with the 
corresponding mass extracted from
lattice simulations. For this we have to construct within the five--dimensional
orbifold lattice theory a {\em gauge invariant} operator which has the proper
symmetries.
This has been discussed in Ref. \cite{Arnold:1995bh} for pure $\SUN$ gauge theories
at nonzero temperature. The Debye mass, which in our context is the Higgs mass,
can be extracted from the exponential fall--off of correlation functions of
gauge invariant operators which are odd under the reflection ${\cal R}$.
The operators proposed in \cite{Arnold:1995bh} can be easily extended to the
orbifold theory and will be studied in forthcoming simulations.

\section{Conclusion}
\label{conclusion}

In this work we constructed non--perturbatively
five--dimensional gauge theories in Euclidean space
with the fifth dimension compactified on the $S^1/\mathbb{Z}_2$ orbifold.

We discussed the possibility of studying these theories on the lattice at
a finite value of the
cutoff $\Lambda=1/a$ given by the inverse lattice spacing and
for energies in the range specified by \eq{range}.
The five--dimensional (four--dimensional) components of the gauge field with
positive ``parity'' under the orbifold projection play the role of the 
Higgs (gluon) field in the dimensionally reduced theory,
defined at the orbifold fixpoints in terms of the Kaluza--Klein zero modes of
these fields.
The ultimate goal of our work is to provide a non--perturbative proof
whether this is a viable field--theoretic scenario, in other words if
a scaling region at finite cutoff exists where the
interactions are appreciable.

We discussed the possible boundary terms localized at the fixpoints of
the orbifold and give a prescription how to derive them. In particular a
non--perturbative proof is given that no boundary term for the Higgs mass
can occur, which would be quadratically divergent in the cutoff $\Lambda$.
We showed that the theories can be formulated in a straightforward way on 
the lattice. Boundary conditions are imposed only for
the links in the four--dimensional boundary planes which belong
to the broken gauge group.
In the naive continuum limit the gauge field propagator implements the
correct Neumann boundary conditions.


\bigskip

{\bf Acknowledgement.}
We are indebted to Martin L\"uscher and Rainer Sommer for precious discussions
and suggestions. We acknowledge helpful discussions with 
Fred Jegerlehner, Theodore Tomaras,
Peter Weisz, Uwe--Jens Wiese, Ulli Wolff and Jean Zinn--Justin.
We are grateful to Burkhard Bunk for his critical reading
of the manuscript. We thank CERN for hospitality during the first stage of this
work.
A special thank goes to the referee for the valuable comments.

\bigskip

\begin{appendix}
\section{Notational conventions}
\label{app_gauge}

The Euclidean gauge action for gauge group $\SUN$ on the manifold $\mathbb{R}^5$
is given by
\bea
 S_5[A] & = & -\frac{1}{2g_0^2}\int{\rm d}^5z\,\tr\{F_{MN}(z)F_{MN}(z)\} \,.
 \label{action5}
\eea
The gauge field $A_M(z)$ belongs to the Lie algebra $\sun$ of $\SUN$
\bea
 A_M^\dagger(z) \,=\, -A_M(z) \,,\quad \tr\{A_M\} \,=\, 0 \,.
\eea
The field strength tensor $F_{MN}(z)$ is defined through
\bea
 F_{MN} & = & [D_M,D_N] =
 \partial_MA_N - \partial_NA_M + [A_M,A_N] \,, \label{fmn}
\eea
where we introduced the gauge covariant derivative
\bea
 D_M & = & \partial_M + A_M \label{cda} \,.
\eea
We denote by $\Omega(z)\in\SUN$ a gauge transformation
in five dimensions. The gauge field transforms as
\bea
 A_M(z) & \Oto & \Omega(z)A_M(z)\Omega(z)^{-1} + \Omega\partial_M\Omega(z)^{-1}
 \,. \label{gta}
\eea
The gauge transformations of the covariant derivative \eq{cda} and the field
strength tensor \eq{fmn} are easily derived
\bea
 D_M & \Oto & \Omega(z)D_M\Omega(z)^{-1} \,, \label{gtcda} \\
 F_{MN}(z) & \Oto & \Omega(z)F_{MN}(z)\Omega(z)^{-1} \,. \label{gtfmn}
\eea
The covariant derivative of the field strength tensor is defined through
\bea
 D_LF_{MN} & = & \partial_LF_{MN} + [A_L,F_{MN}] \label{cdfmn}
\eea
and under gauge transformation it transforms (as its name suggests) like
\bea
 D_LF_{MN}(z) & \Oto & \Omega(z)D_LF_{MN}(z)\Omega(z)^{-1} \,. \label{gtcdfmn}
\eea
The generators $T^A\,,\;A=1,\ldots N^2-1$ of $\SUN$ are typically taken to be
hermitian and traceless. This basis
is spanned for SU(2) by the Pauli $\sigma$-matrices,
for SU(3) by the Gell-Mann $\l$-matrices and so forth.
The generators have the properties
\bea
 [T^A,T^B] = if^{ABC}T^C \,,\quad \tr\{T^AT^B\} = \frac{1}{2}\delta^{AB} \,.
\eea
The connection of this basis with the {\em Cartan--Weyl basis} is
simply to take the Cartan subalgebra, i.e. the commuting generators
$H=\{H_i\}\,,\;i=1,\ldots,N-1$ to be the same.
The remaining generators are combined in pairs of
ladder operators (a raising and a lowering operator)
$E_{\pm\a}\,,\;\a=1,\dots,N(N-1)/2$, which are defined through
(with normalization by $1/\sqrt{2}$)
\bea
E_{\a} \,=\, \frac{1}{\sqrt{2}}(T^{\a(S)}+iT^{\a(A)}) &
\mbox{and}                                            &
E_{-\a} \,=\, \frac{1}{\sqrt{2}}(T^{\a\,(S)}-iT^{\a\,(A)}) \,.\label{laddergen}
\eea
Here by $T^{\a\,(S)}$ we mean the symmetric 
$\SUN$ generator with a 1 in the $mn$--th position
and by $T^{\a\,(A)}$ the anti-symmetric $\SUN$ generator with a $-i$ in the
$mn$--th position ($\a$ labels all possible pairs $mn$, $m\neq n$, with
$m,n=1,\ldots,N-1$).
Each operator $E_\a$ has associated an $(N-1)$--dimensional
vector $\a=\{\a_i\}$ (called root of the operator) such that
\bea
 [H_i,E_{\pm\a}] & = & \pm\a_iE_{\pm\a} \,. \label{laddercomm}
\eea

In the orbifold boundary conditions, the breaking of the gauge symmetry
is realized by a group conjugation with the matrix $g$ defined in \eq{orbma}.
Using the properties \eq{laddercomm} it follows \cite{Hebecker:2001jb}
\bea
 g\,H_i\,g & = & H_i \,, \label{cwgen1} \\
 g\,E_{\pm\a}\,g & = & {\rm e}^{-2\pi i\a\cdot V}E_{\pm\a}
 \hskip .5cm {\rm with} \hskip .5cm
 {\rm e}^{-2\pi i\a\cdot V} \,=\, \pm 1 \,. \label{cwgen2}
\eea
The parity of the generators $\exp(-2\pi i\a\cdot V)$ is determined by the
twist vector $V$. From the relations \eq{laddergen} it follows immediately
that also the hermitian generators $T^A$ have a definite parity $\eta^A$ 
under group conjugation $g\,T^A\,g=\eta^A\,T^A$. We label by $T^a$
the unbroken generators with $\eta^a=1$ and by $T^{\hat{a}}$ the broken
generators with $\eta^{\hat{a}}=-1$. In the adjoint representation for
the generators $T^A$, the matrix $g$ takes the form $g={\rm diag}(\{\eta^a\})$.

\section{Gauge fields with one compact extra dimension}
\label{app_circle}

\subsection{Gauge fields on $S^1$}

When compactifying the fifth dimension on the circle one 
is instructed to define separate gauge fields on overlapping (but not self overlapping)
charts that provide an open cover for the compact space.
The minimum number of such overlapping open sets for $S^1$ is two,
let us call them $O^{(+)}$ and $O^{(-)}$ and their overlaps
$O^{(+-)}_i=(O^{(+)} \cap \, O^{(-)})_i,\; i=1,2$. 
On each open set there is a gauge field that 
under a gauge transformation transforms with its own gauge function
\begin{eqnarray}
&& {\rm on}\; O^{(+)}:\hskip.5cm A_M^{(+)} \longrightarrow \Omega^{(+)} A_M^{(+)} \, {\Omega^{(+)}}^{-1} 
+ \Omega^{(+)}\partial _M {\Omega^{(+)}}^{-1} \\
&& {\rm on}\; O^{(-)}:\hskip.5cm A_M^{(-)} \longrightarrow \Omega^{(-)} A_M^{(-)} \, {\Omega^{(-)}}^{-1} 
+ \Omega^{(-)}\partial _M {\Omega^{(-)}}^{-1} \,.
\end{eqnarray}
One requires that the gauge fields on $O^{(+-)}_i$ (where they are both defined) 
are related by a gauge transformation:
\begin{eqnarray}
&& A_M^{(+)} = G^{(+-)} A_M^{(-)} {G^{(+-)}}^{-1} + 
G^{(+-)}\partial _M {G^{(+-)}}^{-1} \label{rel1}\\
&& A_M^{(-)} = G^{(-+)} A_M^{(+)} {G^{(-+)}}^{-1}+ 
G^{(-+)}\partial _M {G^{(-+)}}^{-1} \,. \label{rel2}
\end{eqnarray}
The $\SUN$--valued functions $G^{(+-)}$ and $G^{(-+)}$ are called
transition functions \cite{Leutwyler:1992yt}
and they are defined on the overlaps of charts $O^{(+-)}_i$.
\eq{rel1} and \eq{rel2} are simultaneously satisfied when the gluing condition  
\begin{equation}
G^{(+-)}G^{(-+)}=\pm 1 \quad \mbox{on $O^{(+-)}_i$} \label{gluing}
\end{equation}
is imposed. Furthermore, their covariance requires that under gauge transformations
they must transform as
\begin{eqnarray}
&& G^{(\pm\mp)} \longrightarrow \Omega^{(\pm)}\; G^{(\pm\mp)}\;  {\Omega^{(\mp)}}^{-1}. \label{Ggt}
\end{eqnarray}
Given the above gauge transformations one can define covariant derivatives
acting on the transition functions such that
\begin{eqnarray}
&& D_M G^{(\pm\mp)} \longrightarrow \Omega^{(\pm)}\; D_M G^{(\pm\mp)}\;  {\Omega^{(\mp)}}^{-1}. 
\end{eqnarray}
This fixes the covariant derivatives to be
\begin{eqnarray}
&& D_M G^{(\pm\mp)} = \partial_M G^{(\pm\mp)} 
+ A_M^{(\pm)}G^{(\pm\mp)} - G^{(\pm\mp)} A_M^{(\mp)}
\end{eqnarray}
(due to \eq{gluing} $D_M G^{(-+)} = (D_M G^{(+-)})^{\dagger}$)
and one can easily see, using eqs. (\ref{rel1}) that
\begin{equation}
D_M G^{(+-)} = D_M G^{(-+)} = 0.
\end{equation}
%

\subsection{Gauge fields on $S^1/Z_2$}

For simplicity we drop in the following the coordinate $x_\mu$ since it is not
affected by the transformations considered.
For an orbifold construction we define the following charts
\begin{eqnarray}
 O^{(+)} \,=\, (-\e ,\pi R + \e) & \mbox{and} &
 O^{(-)} \,=\, (-\pi R - \e ,\e)
\end{eqnarray}
with overlaps $O^{(+-)}_1 = (-\e ,\e )$ and $O^{(+-)}_2 = (\pi R - \e ,\pi R + \e)$
where $0\le \e < (\pi R)/2$.
The coordinates are identified modulo $2\pi R$.

\noindent {\bf Identification under reflection}\\
We introduce the $\mathbb{Z}_2$ transformation
${\cal R}:x_5  \longrightarrow  -x_5$ which maps
${\cal R}\,O^{(\pm)} \,=\, O^{(\mp)}\,$, 
${\cal R}\,O^{(+-)}_i \,=\, O^{(+-)}_i$.
The transformation ${\cal R}$ can be defined also to act on tensor fields defined
on $O^{(\pm)}$ giving as result tensor fields defined on $O^{(\mp)}$.

On the overlaps $O^{(+-)}_i \,,\; i=1,2$ we identify the gauge fields under
the transformation ${\cal R}$ through
\begin{eqnarray}
 {\cal R}\,A_M^{(+)} & = & A_M^{(-)} \,. \label{RidentA}
\end{eqnarray}
This identification is gauge covariant if at the same time the gauge transformations
satisfy on the overlaps
\begin{eqnarray}
 {\cal R}\,\Omega^{(+)} & = & \Omega^{(-)} \,. \label{RidentO}
\end{eqnarray}
Putting together \eq{rel2} with \eq{RidentA}
we obtain the following {\em constraints} for $A_M^{(+)}$
on the overlaps $O^{(+-)}_i \,,\; i=1,2$
\begin{eqnarray}
 {\cal R}\,A_M^{(+)} & = & G^{(-+)} A_M^{(+)} {G^{(-+)}}^{-1}+ 
 G^{(-+)}\partial _M {G^{(-+)}}^{-1} \,.\label{RconstrA}
\end{eqnarray}
Self--consistency of \eq{RconstrA} requires
${\cal R}\,G^{(+-)} \,=\, G^{(-+)}$ and using the gluing condition \eq{gluing}
this gives the constraint
\begin{eqnarray}
 ({\cal R}\,G^{(-+)})\; {G^{(-+)}} & = & \pm 1\,. \label{RidentG}
\end{eqnarray}
From this it follows that
at the {\em fixpoints} $x_5=0$ and $x_5=\pi R$
of the ${\cal R}$ transformation the transition functions satisfy
\begin{eqnarray}
 (G^{(-+)}(0))^2 \,=\, \pm 1 \quad & \mbox{and} & \quad
 (G^{(-+)}(\pi R))^2 \,=\, \pm 1 \,.
\end{eqnarray}

Outside the overlaps we identify further
\begin{eqnarray}
 ({\cal R}\,A_M^{(+)})(x_5) & = & A_M^{(-)}(-x_5)  \,,\quad
 x_5\in[\e,\pi R - \e] \,. \label{Routside}
\end{eqnarray}
Therefore we can set up the gauge theory on the one chart $O^{(+)}$
with gauge field $A_M\equiv A_M^{(+)}$ and a {\em spurion field} defined
on the overlaps $O^{(+-)}_i \,,\; i=1,2$ through
\begin{eqnarray}
 {\cal G} \,=\, G^{(-+)} \quad &,& \quad ({\cal R}\,{\cal G})\,{\cal G} \,=\, \pm 1
 \label{app_spurion}
\end{eqnarray}
which, using \eq{Ggt} and \eq{RidentO} has the gauge transformation
under $\Omega\equiv\Omega^{(+)}$
\begin{eqnarray}
 && {\cal G} \longrightarrow ({\cal R}\,\Omega)\,{\cal G}\,\Omega^{-1} \,.
 \label{Ggt2}
\end{eqnarray}

\end{appendix}

\bibliography{orbi}           
\bibliographystyle{h-elsevier.bst}   

\end{document}